\documentclass{webofc}
\usepackage[varg]{txfonts}

\usepackage{hyperref}

\usepackage{listings}
\usepackage{xcolor}
\begin{document}
\title{A Portable Implementation of RANLUX++}

\author{
  \firstname{Jonas} \lastname{Hahnfeld}\inst{1}\fnsep\thanks{\email{jonas.hahnfeld@cern.ch}}
  \and
  \firstname{Lorenzo} \lastname{Moneta}\inst{1}
}

\institute{CERN, Geneva, Switzerland}

\abstract{High energy physics has a constant demand for random number generators (RNGs) with high statistical quality.
In this paper, we present ROOT's implementation of the RANLUX++ generator.
We discuss the choice of relying only on standard C++ for portability reasons.
Building on an initial implementation, we describe a set of optimizations to increase generator speed.
This allows to reach performance very close to the original assembler version.
We test our implementation on an Apple M1 and Nvidia GPUs to demonstrate the advantages of portable code.
}

\maketitle

\section{Introduction}
\label{sec:introduction}

Over the past decades, researchers have developed many generators to produce streams of pseudo-random numbers.
However, only few of them provide the excellent quality needed for many applications in high energy physics~\cite{review}.
In addition to the family of MIXMAX~\cite{mixmax}, this includes the subtract-with-borrow generator RANLUX~\cite{ranlux}.
The latter has proven theoretical foundations~\cite{luscher}, but requires skipping generated numbers, which is expensive.

RANLUX++ is an extension of RANLUX using the fact that it is equivalent to a Linear Congruential Generator (LCG) with a very large multiplier~\cite{sibidanov}.
This allows to advance the state without computing unneeded intermediate results by exploiting the modular arithmetic.
Furthermore, it is possible to efficiently skip a sequence of random numbers and seed the generator~\cite{sibidanov}.

For performance reasons, the core routines of the original version are written in assembler for the x86 architecture.
This ties the implementation to a single platform, which is not suitable for inclusion into the ROOT data analysis framework~\cite{root}.
The development of specialized implementations for every supported architecture would be a large effort and maintenance cost.
Moreover, it would be necessary to repeat this difficult task whenever a new platform has to be supported.
For these reasons, we propose a portable implementation of RANLUX++ written in standard C++.
This allows to use the compiler as a means of abstraction for targeting many architectures at once.

The remainder of this paper is structured as follows:
First we provide details about the implementation in Section~\ref{sec:implementation}.
In Section~\ref{sec:optimization}, we present optimizations for x86, our initial target.
This includes a comparison with the original version written in assembler.
Section~\ref{sec:portability} discusses the aspect of portability.
Finally we draw conclusions in Section~\ref{sec:conclusion}.

\section{Portable RANLUX++ for ROOT}
\label{sec:implementation}

In our first version of the portable implementation, each instance of the generator stored the LCG state.
Because there is no portable type for numbers of the required size, we use \texttt{uint64\_t[9]}\footnote{
We use the notation \texttt{type[n]} to denote an array with \texttt{n} elements of type \texttt{type}.}
to store them in 9 separate 64 bit numbers for a total of 576 bits.
This was proposed by Sibidanov~\cite{sibidanov} and is also used in the assembler implementation.
Furthermore, the generator needs to remember which bits to use for the next random number.
For this, we store the current position in the state as \texttt{int}.

However, drawing bits from the LCG state leads to a bias in the generated numbers as further explained in Section~\ref{sec:floating-point}.
For this reason, our final version of the generator stores RANLUX numbers and converts to the LCG state as required.
To reduce the required memory, the RANLUX numbers are packed as 576 bits in a member of type \texttt{uint64\_t[9]}.
Additionally, the generator also needs to store the carry bit of the RANLUX state~\cite{ranlux}.

\subsection{Multiplication of two 576~bit numbers}
\label{sec:multiplication}

For the LCG operation, we have to implement our own arithmetic operations on its representation.
The first one is \texttt{multiply9x9} to multiply two 576~bit numbers stored as \texttt{uint64\_t[9]}.
Their product is returned as \texttt{uint64\_t[18]} which is computed as follows:
For every 64~bit in the output, there are up to 9~contributing pairs of 64~bit numbers from the input.
The code multiplies each of these pairs and sums their individual products.
The lower 64~bits of the sum constitutes one entry of the output, while the rest is propagated to the next entry.

\subsection{Modulo operation}
\label{sec:modulo}

The second operation of an LCG requires to determine the remainder of a division by a modulus.
For RANLUX, this modulus is known to have the value $m = 2^{576} - 2^{240} + 1$.
Taking advantage of the structure, Sibidanov notes that it is possible to implement the operation using additions, subtractions, and bit shifts~\cite{sibidanov}.
We follow the described algorithm in our function~\texttt{mod\_m}, with an additional fix to ensure that the result is smaller than $m$ in all cases.

\subsection{Generating floating point numbers}
\label{sec:floating-point}

The original version of RANLUX++ treats the LCG state as entropy pool~\cite{sibidanov}.
Many applications, however, require floating point number that are uniformly distributed in the interval $[0, 1)$.
For the conversion, Sibidanov uses 52 random bits to set the mantissa of a \texttt{double}.
However, this leads to a bias in the generated numbers because the modulus $m$ is not a power of two.
For this reason, the upper bits of the LCG state have a higher probability of being 0.

To avoid possible problems, our version instead converts the LCG state back to RANLUX numbers, that are equally distributed\footnote{
Similar functionality is also available as an optional interface in the assembler version, but the paper does not describe its implementation details.}.
Once all converted numbers got used, the generator advances its temporary LCG state to produce the next set of RANLUX numbers.
Additionally, the final version uses only 48 bits for each floating point value.
This reinforces the connection to the theoretical properties derived from dynamical chaotic systems~\cite{luscher}.

\subsection{Integration with ROOT}
\label{sec:root}

As interface to developers, we expose the class \texttt{RanluxppEngine}.
It has functions to seed the generator and skip random numbers without generating them.
Both follow the description by Sibidanov and use fast exponentiation by squaring~\cite{sibidanov}.
To generate random numbers, the class provides \texttt{IntRndm()} for integer numbers and \texttt{Rndm()} for double precision.
The latter is a virtual method and implements the interface defined by \texttt{TRandomEngine}.
This allows to instantiate the template \texttt{RandomFunctions} to generate random numbers of certain distributions, for example Gaussian.

\subsection{Results of empirical tests}
\label{sec:empirical}

% To test our first implementation, we compare the generated numbers to the assembler version by Sibidanov~\cite{sibidanov}.
% We use the same initialization method of skipping $2^{96} \cdot seed$ states to obtain the same sequences.
% Additionally the version by Sibidanov advances the state once after initialization.\footnote{
% This behavior is not documented in the paper and might be a result of implementation details.}
% When skipping the first 11 numbers in our implementation, we observe the same random numbers as the assembler version by Sibidanov.

To test our implementation, we subject the generated numbers to empirical statistical testing:
\emph{TestU01} is a library of utilities for RNGs developed by L'Ecuyer and Simard~\cite{TestU01}.
The library contains implementations of generators, statistical tests, and predefined ``batteries'' of tests.
The most stringent set of tests for uniform random numbers in the interval $[0, 1)$ is called \texttt{BigCrush}.
It performs a total of 106 tests and ran for around 3 hours (including the optimizations described in the following section) with the following summary:
\begin{lstlisting}
========= Summary results of BigCrush =========

 Version:          TestU01 1.2.3
 Generator:        RanluxppEngine2048
 Number of statistics:  160
 Total CPU time:   03:17:15.86

 All tests were passed
\end{lstlisting}

Due to their statistical nature, some tests may warn when starting from certain states.
For example, the test \texttt{MaxOft} reports a p-value of $0.9990$ when running both \texttt{SmallCrush} and then \texttt{BigCrush} with the default seed.\footnote{Above output was obtained with seed $2718$.}
This value is the upper limit of the interval $[0.001, 0.9990]$ that is considered normal by TestU01.
As the tests pass with other seeds, we consider the warning to be spurious.

\section{Optimization on x86}
\label{sec:optimization}

After finishing the first implementation, this section looks at performance on x86 and discusses optimizations.
For a fair comparison with the assembler version by Sibidanov, we give numbers for the initial generator without conversion to RANLUX numbers.
On a recent system with an AMD~Ryzen~9~3900 running CentOS~8, we use the default GNU compiler in version~8.3.1.
When building ROOT in \texttt{Release} mode with optimizations, it takes an average of 28.5\,ns to generate one random number.
This is around 3.26 times slower than the assembler version at 8.74\,ns per number.

\begin{figure}
\centering
\includegraphics[width=\textwidth]{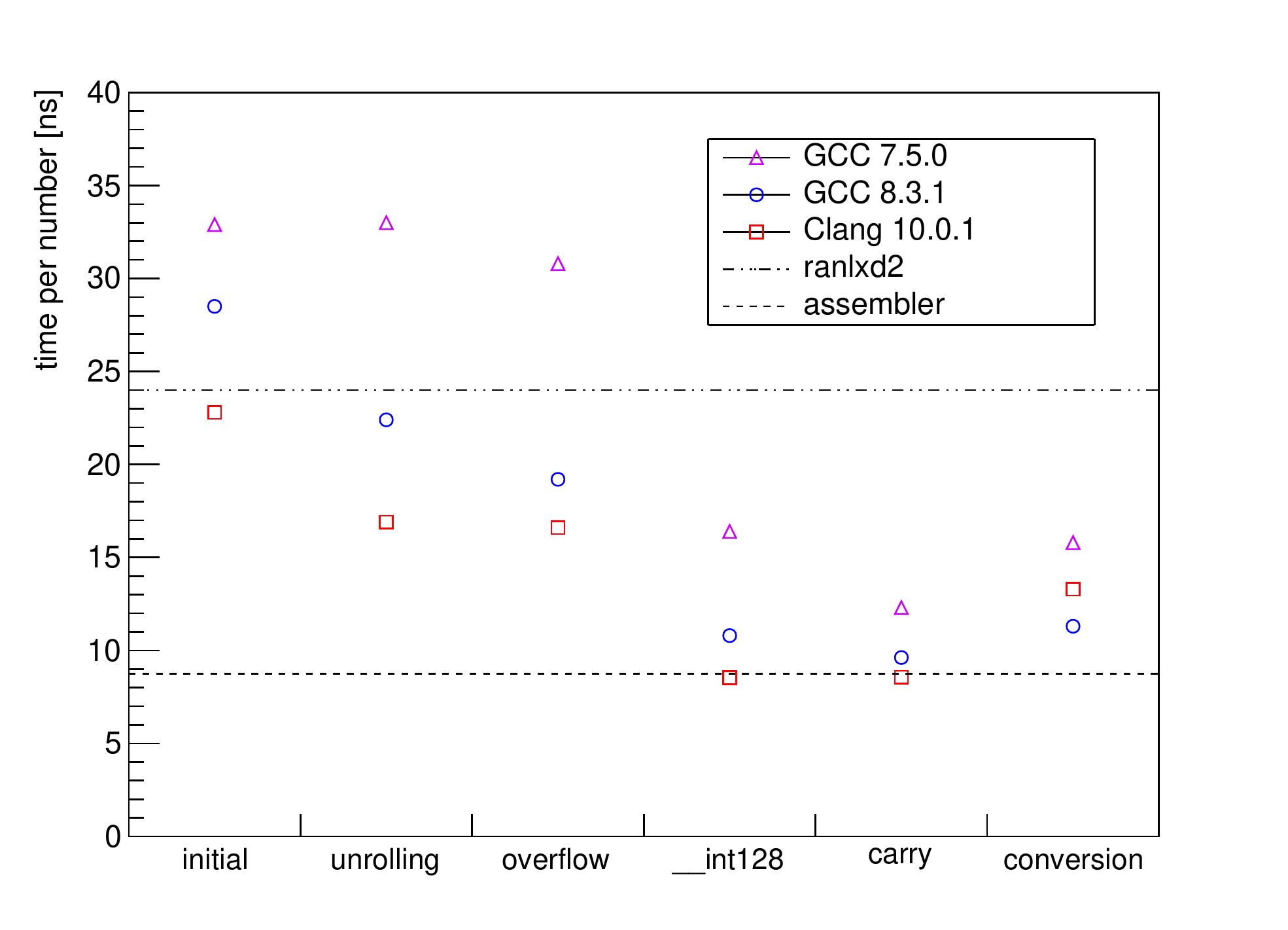}
\caption{Performance results after each of the presented optimizations with different compilers.}
\label{fig:optimization}
\end{figure}

Profiling shows that the larger portion of the runtime is spent for the multiplication of two 576 bit numbers as described in Section~\ref{sec:multiplication}.
As a consequence, most of the optimizations presented in the following focus on the function \texttt{multiply9x9}.
Finally the optimization of the carry propagation in Section~\ref{sec:carry-propagation} also applies to \texttt{mod\_m}.
Performance results after each optimization on the same system are presented in Figure~\ref{fig:optimization}.
In addition to GCC 8.3.1, we also measure with Clang~10.0.1 as offered by CentOS~8 and the older GCC~7.5.0.
The latter is the same major version as the default compiler found in Ubuntu~18.04~LTS.
The final column of Figure~\ref{fig:optimization} shows a measurement of the final version including the conversion to RANLUX numbers.
The horizontal lines show the measured runtime for the assembler version of RANLUX++ by Sibidanov, and of \texttt{ranlxd2} written by Lüscher~\cite{ranlxd2}.

\subsection{Loop unrolling}
\label{sec:loop-unrolling}

The portable code relies on loops to implement the multiplication and modulo operation in software.
This induces overhead when evaluating the exit condition and incrementing the loop variable.
Furthermore, the loop structure leads to jumps in the code that hinder instruction level parallelism.
Fortunately loop unrolling is a well known compiler optimization to tackle this problem:
It duplicates the loop body to enable other optimization passes and generate more efficient code.
In case of constant loop bounds, it is even possible to fully unroll the loop.

One disadvantage of loop unrolling is the increased code size.
For this reason, compilers are conservative in their unrolling heuristics.
During experiments, we found that unrolling the loops in \texttt{multiply9x9} consistently improves performance.
By adding \texttt{\#pragma} directives as shown in Listing~\ref{lst:loop-unrolling}, it is possible to help the compiler.
As can be seen in Figure~\ref{fig:optimization}, this improves performance for GCC 8.3.1 and Clang 10.0.1 by up to 25\,\%.
GCC 7.5.0 does not benefit from the change because \texttt{\#pragma GCC unroll} was only added in GCC 8.

\begin{lstlisting}[float, caption={Unrolling the outer loop in \texttt{multiply9x9} for multiplying two 576 bit numbers. Another inner loop is annotated similarly.}, label=lst:loop-unrolling]
#if defined(__clang__) || defined(__INTEL_COMPILER) \
                       || defined(__CUDACC__)
#pragma unroll
#elif defined(__GNUC__) && __GNUC__ >= 8
// This pragma was introduced in GCC version 8.
#pragma GCC unroll 18
#endif
   for (int i = 0; i < 18; i++) {
      // ...
   }
\end{lstlisting}

\subsection{Overflow handling}
\label{sec:overflow-handling}

For the multiplication of two 576 bit numbers, the portable code is constrained by the available data types:
It can at most multiply two 32 bit values at a time to stay within the 64 bits of \texttt{uint64\_t}.
This makes it necessary to handle overflows when summing up the partial results.
By arranging the operations differently, it is possible to avoid conditional execution.
Together with full unrolling, this eliminates all jumps from the generated machine code for \texttt{multiply9x9} with GCC 8.3.1.
During execution, this leads to performance improvements of up to 15\,\% for the two versions of GCC.
There is no change for Clang because its optimizations already transformed the original code.

\subsection{Native multiplication}
\label{sec:native-mult}

An even faster way is to let hardware handle the multiplication of two 64 bit numbers.
This is possible with the extension type \texttt{\_\_int128} supported by both GCC and Clang.
Using this type, it is possible to multiply 64 bit numbers with 128 bit precision.
For that, the compiler can make use of special instructions available for the targeted hardware.
Afterwards the upper and lower halves can be extracted as 64 bit numbers as shown in Listing~\ref{lst:native-mult}.
Figure~\ref{fig:optimization} shows that taking advantage of \texttt{\_\_int128} improves performance of all three compilers by more than 40\,\%.

\begin{lstlisting}[float=b, caption={Multiplying the two 64 bit numbers \texttt{fac1} and \texttt{fac2} using the type \texttt{\_\_int128}. The code is only compiled if the extension is available and can be disabled explicitly to test the fallback.}, label=lst:native-mult]
unsigned __int128 prod = fac1;
prod = prod * fac2;

uint64_t upper = prod >> 64;
uint64_t lower = static_cast<uint64_t>(prod);
\end{lstlisting}

\subsection{Carry propagation}
\label{sec:carry-propagation}

Additional tests showed inferior performance with older compilers.
Part of this is related to missing loop unrolling because the \texttt{\#pragma} is not available as mentioned in Section~\ref{sec:loop-unrolling}.
An additional problem is the generation of jumping code to propagate carry bits in case of overflows.
Instead of relying on inline assembly code, we present a portable solution after refactoring the common code:
Listing~\ref{lst:carry-prop-if} shows the initial implementation to detect an overflow and propagate the carry.
It uses an \texttt{if} statement and the tested versions of GCC emit jumping code.
In contrast, Listing~\ref{lst:carry-prop-add} performs the addition unconditionally.
This modification results in good code for all tested compilers.
For the older GCC 7.5.0 in particular, it improves performance by 25\,\%.

\begin{figure}
\centering
\begin{minipage}[t]{0.45\textwidth}
\begin{lstlisting}[caption={Carry propagation using an \texttt{if} statement.}, label=lst:carry-prop-if]
uint64_t add = a + b;
if (add < a)
   carry++;
\end{lstlisting}
\end{minipage}
\quad
\begin{minipage}[t]{0.45\textwidth}
\begin{lstlisting}[caption={Carry propagation by adding the overflow boolean value.}, label=lst:carry-prop-add]
uint64_t add = a + b;
// Add either 0 or 1
carry += (add < a);
\end{lstlisting}
\end{minipage}
\end{figure}

\subsection{Performance summary}
\label{sec:perf-summary}

Using all optimizations, the performance of the portable code gets competitive to the assembler version by Sibidanov and is sometimes able to beat it:
When compiled with Clang~10.0.1, it takes an average of 8.56\,ns per number compared to 8.74\,ns with the assembler code.
This is closely followed by GCC~8.3.1 at 9.62\,ns, which is a factor of 3x faster than the initial version.
The older GCC~7.5.0 is slower at 12.3\,ns, which is due to the missing unrolling directives mentioned before.
The conversion to and from RANLUX numbers slows the generator down by~17\,\% to~55\,\%, depending on the compiler.
However, the generator is still a factor of 2x faster than the version by Lüscher using the original algorithm.

\section{Portability}
\label{sec:portability}

After optimizing performance on x86, the following sections discuss the portability aspect of our implementation.
For this, we first repeat the previous measurements on a Mac mini with Apple's M1 processor based on the ARM architecture.
Afterwards, we present the throughput of random numbers on an Nvidia GPU and compare it with cuRAND.

\subsection{Apple M1}
\label{sec:apple-m1}

\begin{figure}
\centering
\includegraphics[width=\textwidth]{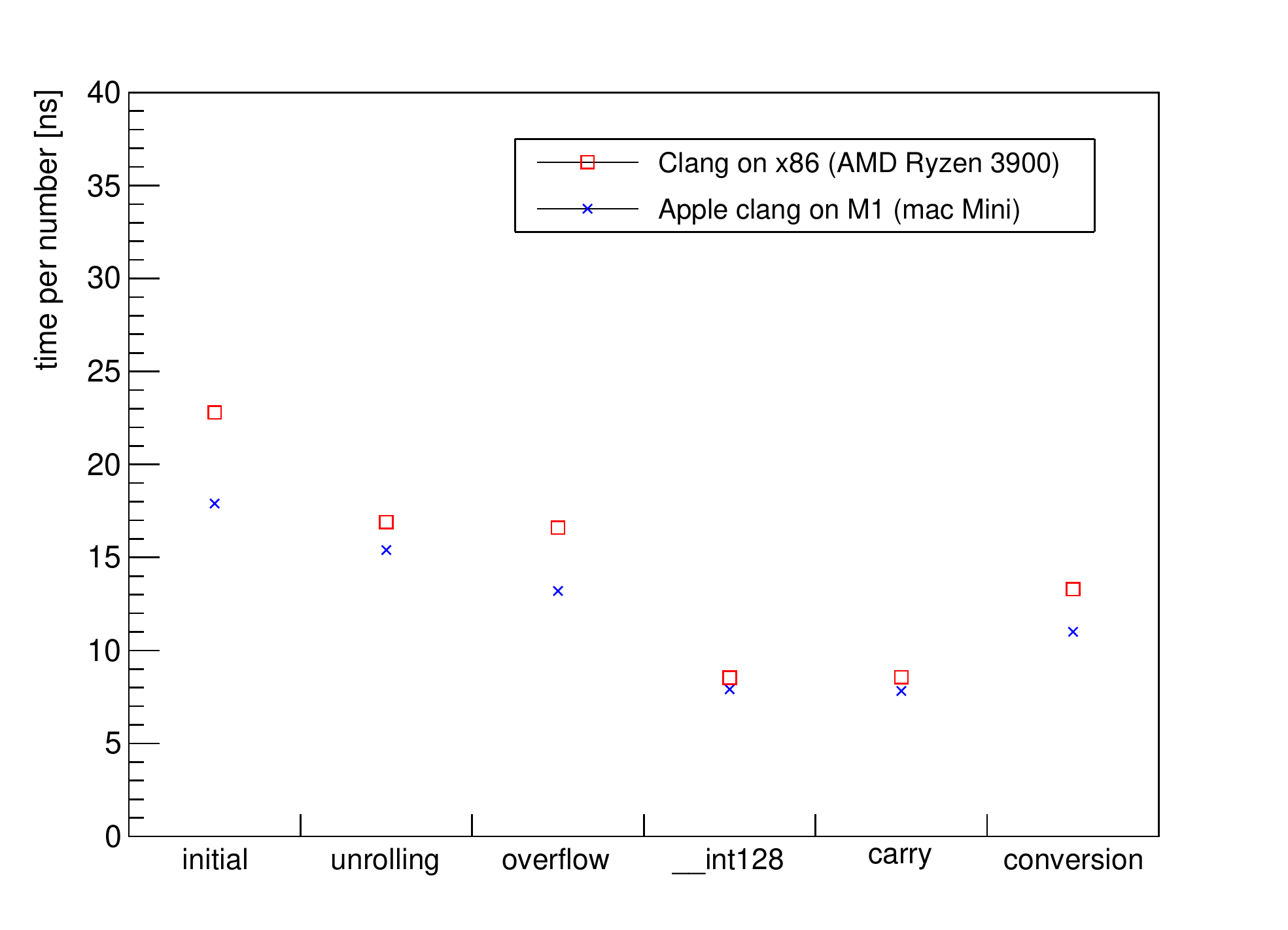}
\caption{Performance results on Apple's M1 processor compared to Clang on x86.}
\label{fig:apple-m1}
\end{figure}

On the Mac mini, we compile ROOT in \texttt{Release} mode with Apple clang version~12.0.5 \texttt{(clang-1205.0.22.9)}.
Figure~\ref{fig:apple-m1} shows the time per number for the optimization steps described in the previous section.
As before, the first five measurements exclude the conversion to RANLUX numbers, which is only shown in the last column.
It can be seen that the optimizations lead to portable performance improvements on this new platform.
In this particular benchmark, single-core performance is also consistently better on the Apple M1 when compared to an AMD~Ryzen~3900.

Furthermore, it is important to note that our implementation yields the same sequence of random numbers given the same seed.
This means analysis results will be reproducible no matter if run on x86 or one of Apple's new M1 processors.

\subsection{Nvidia GPUs}
\label{sec:nvidia-gpus}

With the code written in standard C++, it is possible to reuse most of the implementation on Nvidia GPUs with CUDA.
Only minor changes are needed for the interfaces, where we remove the dependency on ROOT by not inheriting from \texttt{TRandomEngine}.
Furthermore, we hardcode the luxury level \texttt{p = 2048} as the recommended value by Sibidanov~\cite{sibidanov}.
Finally, we add the annotations \texttt{\_\_host\_\_ \_\_device\_\_} to the functions for generating a random number as well as seeding and advancing the state.
In particular, this transitively includes \texttt{multiply9x9} and \texttt{mod\_m} to perform the arithmetic operations and the functions to convert to and from RANLUX numbers.

The code located inside \texttt{multiply9x9} can be copied with only one change:
The type \texttt{\_\_int128} described in Section~\ref{sec:native-mult} is not supported on the GPU.
However, it continues to be available on the host, so it needs to be disabled only on the device.
This is achieved by checking \texttt{!defined(\_\_CUDA\_ARCH\_\_)} with the preprocessor.
No change is needed for \texttt{mod\_m} and the conversion to and from RANLUX numbers.

To test RANLUX++ on a GeForce RTX 2070 SUPER, we use 1024~blocks of 128~threads each.
During setup, we initialize one state per thread to avoid synchronization.
Then we measure how long it takes all threads to generate 1~million \texttt{double} values each.
To avoid the compiler removing the generation loop, we sum all numbers per thread and store them into global memory at the end.

We find that RANLUX++ on this GPU takes around 11.8~seconds to generate a total of $1.3 \cdot 10^{11}$ random numbers.
For comparison, the default XORWOW generator from cuRAND needs around 3.1~seconds, but fails a number of statistical tests.
However, performance of RANLUX++ gets significantly worse when threads need to advance their states at different times.
To provoke this situation, we make the first thread skip zero random numbers at the beginning, the second thread one number, and so on.
This leads to significant thread divergence because the generators need to perform the LCG operation for every twelfth number.
In this setup, we measure more than 80~seconds for $1.3 \cdot 10^{11}$ random numbers.
This effect is not visible with XORWOW because its generation time is uniform for all drawings.

As another important feature, the state can be copied between host and device with a simple \texttt{cudaMemcpy}.
This allows to move control flow between the two while continuing the same sequence of random numbers.
To the best of our knowledge, this is not possible with cuRAND at the time of writing.

\section{Conclusion}
\label{sec:conclusion}

We presented a portable implementation of the RANLUX++ generator, an LCG extension of RANLUX.
The discussed implementation relies only on standard C++ and was integrated with the ROOT data analysis framework.
We showed that a set of portable optimizations leads to performance close to the previously available assembler version.
Finally, we discussed execution on Apple's M1 processor and throughput on Nvidia GPUs.
Due to its proven statistical properties, we envision RANLUX++ to be used in a wide range of applications.
This includes Monte-Carlo simulations, where especially the portability to GPUs might become important.

\section*{Acknowledgements}

The authors would like to thank Martin Lüscher for reporting the bias described in Section~\ref{sec:floating-point}.
We also gratefully acknowledge his help and advice on implementing the conversion to RANLUX numbers.

\bibliography{paper}

\end{document}